\documentclass[12pt,english]{iopart}
\newcommand{\be}{\begin{equation}}
\newcommand{\ee}{\end{equation}}
\newcommand{\bea}{\begin{eqnarray}}
\newcommand{\eea}{\end{eqnarray}}

\newcommand{\bJ}{{\bf J}}
\newcommand{\bj}{{\bf j}}
\newcommand{\bE}{{\bf E}}

\newcommand{\ket}[1]{\left|#1\right>}
\newcommand{\bra}[1]{\left<#1\right|}

\newcommand{\var}{{\rm var}}

\usepackage{iopams}
\usepackage[T1]{fontenc}
\usepackage{hyperref}
\usepackage{graphicx,color}
\usepackage{bm}
\usepackage[utf8]{inputenc}
\usepackage{upgreek}
\usepackage{verbatim}
\usepackage{babel,url}
\usepackage{ulem}
\makeatother

\newcommand{\mtext}[1]{#1}
\newcommand{\otext}[1]{{\color{green}}}

\newcommand{\ntext}[1]{{\color{black}#1}}

\begin{document}
\title{Nonlinear metrology with a quantum interface}
\author{M. Napolitano\(^*\), M. W. Mitchell}

\address{ICFO-Institut de Ciencies Fotoniques, 08860 Castelldefels (Barcelona),
Spain}
\ead{\(^*\)mario.napolitano@icfo.es}

\begin{abstract}
We describe nonlinear quantum atom-light interfaces and nonlinear quantum metrology in the collective continuous variable formalism.  We develop a nonlinear effective Hamiltonian in terms of spin and polarization collective variables and show that model Hamiltonians of interest for nonlinear quantum metrology can be produced in $^{87}$Rb ensembles.  With these Hamiltonians, metrologically relevant atomic properties, e.g. the collective spin, can be measured better than the ``Heisenberg limit'' $\propto 1/N$.  In contrast to other proposed nonlinear metrology systems, the atom-light interface allows both linear and non-linear estimation of the same atomic quantities.

\submitto{New Journal of Physics}

\end{abstract}



\maketitle

\newcommand{\myvec}[1]{{\bf #1}}
\newcommand{\myvecE}[1]{\pmb{#1}}
\newcommand{\FA}{{\mathcal E}}
\newcommand{\PE}{{\hat{S}}}
\newcommand{\upin }{^{(\rm in)}}
\newcommand{\upout }{^{(\rm out)}}
\newcommand{\PRLsection}[1]{\noindent{\it #1}---}

\label{sec:intro}

\newcommand{\envelope}{\FA}
\newcommand{\UpSym}{{\uparrow}}
\newcommand{\DownSym}{{\downarrow}}
\newcommand{\DownwardDipole}{\myvec{d}_\DownSym}
\newcommand{\UpwardDipole}{\myvec{d}_\UpSym}
\newcommand{\HFSSym}{\Delta}
\newcommand{\PolUnitSym}{\myvec{e}}

\newcommand{\bF}{{\bf F}}
\newcommand{\bff}{{\bf f}}
\newcommand{\bS}{{\bf S}}

\newcommand{\Sx}{{S_x}}
\newcommand{\Sy}{{S_y}}
\newcommand{\Sz}{{S_z}}
\newcommand{\Jx}{{J_x}}
\newcommand{\Jy}{{J_y}}
\newcommand{\Jz}{{J_z}}
\newcommand{\fx}{{f_x}}
\newcommand{\fy}{{f_y}}
\newcommand{\fz}{{f_z}}

\newcommand{\PAble}{\stackrel{\leftrightarrow}{\alpha}}
\newcommand{\Al}{\FA}
\newcommand{\plus}{+}
\newcommand{\minus}{-}

\newcommand{\myexpect}[1]{\left<\right.#1\left.\right>}

\section{Introduction}

In quantum metrology, a quantum state is prepared, 
evolves under the action of a Hamiltonian containing a parameter $x$
of interest, and is measured. The parameter is estimated
from the measurement outcome and knowledge of the system and
Hamiltonian. In most problems, the Hamiltonian is assumed to act
in the same way on each of $N$ systems (e.g. atoms), and  precision
scales as $\delta x \propto N^{-1/2}$ for product states, and 
down to  $\delta x \propto N^{-1}$ for
entangled states (``Heisenberg limit'' scaling)
\cite{Giovannetti2006}. A number of studies
\cite{Boixo2007,Boixo2008PRA,Boixo2008PRL} have considered also
nonlinear quantum metrology, in which the Hamiltonian describes a 
$k$-system coupling with strength $x'$. 
Remarkably, the
scaling is $\delta x' \propto N^{-k+1/2}$ or $\delta x'
\propto N^{-k}$ for independent or entangled states, respectively
\cite{Boixo2007}. Because this improves upon the best possible scaling for
the linear case, it has been called ``Super-Heisenberg'' (SH) scaling 
\cite{Boixo2008PRL}.  Proposed implementations include scattering in 
Bose condensates
\cite{Boixo2008PRL}, Duffing nonlinearity in nano-mechanical
resonators \cite{WoolleyNJP2008}, two-pass effective
nonlinearity with an atomic ensemble \cite{Chase2009}, and
Kerr nonlinearities \cite{Luis2004, Beltran2005, Luis2007}.  

Here we describe nonlinear metrology applied to measurement of collective spin variables of atomic ensembles.  Atomic ensembles with long coherence-time internal degrees of freedom, e.g. nuclear spin, are essential elements of many quantum information and quantum metrology protocols including quantum memory \cite{PolzikMemory}, quantum non-demolition measurement \cite{Koschorreck2010}, spin squeezing \cite{Windpassinger2009,SchleierSmith2010,Takano2009}, and magnetometry \cite{RomalisNP2007}.

We follow the approach of collective continuous variables (CCV), in which both light and atoms are described by macroscopic quantum variables.  
In the case of spinor atoms interacting with polarized light, the $N_A$ atoms are described by the collective spin $\bF \equiv \sum_i \bff^{(i)}$ where $\bff^{(i)}$ is the spin of the $i$-th atom.  $\bF$ obeys the commutation relations $[F_x,F_y] = i \hbar F_z$ and cyclic permutations, and can itself be considered a macroscopic spin variable.  The light is described by its electric field ${\bf E} = \myvecE{\FA} + \myvecE{\FA}^*$, where $\myvecE{\FA}$ is the positive-frequency part.   The Stokes vector ${\bf S}$ with components $S_i =  (\FA_+^*,\FA_-^*)\sigma_i (\FA_+,\FA_-)^{\rm{T}} $ where the subscript indicates plus/minus circular polarization, $\sigma_i$ are the Pauli matrices and $\sigma_0$ is the identity. 
As described by several authors \cite{Kupriyanov2005, Geremia2006, deEchaniz2008}, the electric dipole interaction $h_{\rm int} = - \bE\cdot \myvec{d}$, taken in second order perturbation theory, gives rise to an effective (single-atom) Hamiltonian of the form
\bea
h_{\rm eff} &=& \sum_{k} \frac{\myvecE{\FA}^* \cdot \myvec{d}_\DownSym \ket{\phi_k} \bra{\phi_k} \myvec{d}_\UpSym \cdot \myvecE{\FA}}{\hbar \delta_k} \\
& = & \myvecE{\FA}^* \cdot \PAble \cdot \myvecE{\FA}
\eea
where 
 $d_\UpSym,d_\DownSym$ are the parts of the dipole operator causing upward/downward transitions, $\delta_k$ is the detuning from resonance of the $k$th state, and $\PAble$ is the tensor polarizability operator.   The effective Hamiltonian for the ensemble $H_{\rm eff} = \sum_i h_{\rm eff}^{(i)}$ can then be decomposed into irreducible tensor components as
\bea
H_{\rm eff}^{(2)}  & = &  \alpha_{1}\Sz\Jz+\alpha_{2}\left(\Sx\Jx+\Sy\Jy\right) \,\,,\label{eq:HEff2}\
\eea
plus terms in $S_0$ which do not interact with the optical polarization.  Here $\alpha_{1,2}$ describe the vectorial, and tensorial components of the interaction, respectively, and the atomic collective variable is $\bJ \equiv \sum_i \bj^{(i)}$ with $j_x \equiv \left(\fx^{2}-\fy^{2}\right)/2$, $j_y \equiv \left(\fx \fy+\fy \fx\right)/2$, $j_z \equiv f_z/2$ and $j_0 \equiv f_z^2/2$.  The ratio of the $\alpha_i$ can be tuned by adjusting the optical frequency $\omega$, giving a variety of Hamiltonians interesting for quantum information tasks \cite{deEchaniz2008}.

To apply this formalism to nonlinear metrology, we generalize the CCV method to the nonlinear optics regime, i.e., we include higher-order processes in the effective Hamiltonian. For this purpose, na\"{i}ve application of higher-order perturbation theory fails due to the appearance of vanishing resonance denominators, and degenerate perturbation theory \cite{Klein1974} is required.  We present the method by way of an example, the $D_2$ line of $^{87}$Rb, one of the most used transitions for atom-light interactions.

\section{Derivation of the Effective Hamiltonian}

We consider the $5 ^{2}S_{1/2} \rightarrow 5 ^{2} P_{3/2}$
transition (the $D_2$ transition at 780 nm).  The ground states
are $\ket{F,m_F}$ with $(F,m_F) = (1,-1), (1,0), (1,1), (2,-2),
\ldots, (2,2)$.  The excited states are $\ket{F',m_{F'}}$ with
$(F',m_{F'}) = (0,0), (1,-1),  (1,0),  \ldots, (3,3)$.  We use
these states as a basis, with the ground states preceding the
excited states. We calculate the single-atom Hamiltonian $h_{\rm
eff}$, and note that the ensemble Hamiltonian $H_{\rm eff} =
\sum_i h_{\rm eff}^{(i)}$ is found simply by replacing single-atom
operators such as $\bf j$ with collective operators ${\bf J}\equiv
\sum_i {\bf j}^{(i)}$. The unperturbed Hamiltonian is $h_{0} =
\hbar \sum_l \omega_l \ket{l}\bra{l}$, or in matrix notation
\begin{eqnarray}
h_{0} &=&  \hbar (\omega_{F=1} I_3 \oplus \omega_{F=2} I_5
\oplus \omega_{F'=0} I_1 \oplus \omega_{F'=1} I_3 \nonumber \\ & &
\oplus \omega_{F'=2} I_5   \oplus \omega_{F'=3} I_7).
\end{eqnarray}
where $\oplus$ indicates a direct sum, and $I_d$ is the identity
matrix of dimension $d$.   We choose the origin of energy such
that $\omega_{F=1}=0$, and define $\HFSSym \equiv \omega_{F=2}$.
We work in a frame rotating with the laser frequency $\omega =
\omega_{F'=0} + \delta$.  In this frame, the Hamiltonian is
\begin{eqnarray}
h_{0} &=& \hbar( 0 I_3 \oplus \HFSSym I_5 \oplus \delta_0 I_1
\oplus \delta_1 I_3 \oplus \delta_2 I_5   \oplus \delta_3 I_7)
\end{eqnarray}
where $\delta_{F'} \equiv \omega_{F'}-\omega_{}$.

In the rotating wave approximation, the single-atom perturbation 
$v = h_{\rm int} = - {{\bf E}}\cdot{\bf d}$ is approximated as $v \approx
\myvecE{\FA} \cdot \UpwardDipole  + \myvecE{\FA}^* \cdot
\DownwardDipole$.  If $\Al_{\pm}$ are the amplitudes
for the sigma-plus/minus components, respectively, of
$\myvecE{\FA}$, then \begin{equation} \bra{F',m_{F'}} V
\ket{F,m_F} = \Al_q \bra{F',m_{F'}} e r_q \ket{F,m_F}
\end{equation}
with $q = m_{F'}-m_F$.  Note that $q=0$ transitions
($\pi$-transitions) are not considered because the $z$-propagating
beam cannot contain this polarization. The dipole matrix elements
are related to the ``matrix element'' $\bra{J}|e r_q |\ket{J'}
\equiv D_{JJ'} \approx 3.584 10^{-29} {\rm C \cdot m}$ by
angular-momentum addition rules.  We follow the conventions given
in Steck \cite{Steck2009}.
%
In this way, we arrive to the perturbation Hamiltonian
\begin{equation}
V = \left(
\begin{array}{cc}
0_8 & V_\UpSym^\dagger \\
V_\UpSym & 0_{16}
\end{array}
\right)
\end{equation}
where $0_d = 0 I_d$ and $V_\UpSym\equiv\sqrt{5} D_{JJ'} \times$
\begin{equation}
 \left(
\begin{array}{cccccccc}
 \frac{\Al_{\plus}}{\sqrt{30}} & 0 & \frac{\Al_{\minus}}{\sqrt{30}} & 0 & 0 & 0 & 0 & 0 \\
 0 & \frac{\Al_{\minus}}{2 \sqrt{6}} & 0 & \frac{\Al_{\plus}}{10} & 0 & \frac{\Al_{\minus}}{10 \sqrt{6}} & 0 & 0 \\
 \frac{\Al_{\plus}}{2 \sqrt{6}} & 0 & \frac{\Al_{\minus}}{2 \sqrt{6}} & 0 & \frac{\Al_{\plus}}{10 \sqrt{2}} & 0 &
   \frac{\Al_{\minus}}{10 \sqrt{2}} & 0 \\
 0 & \frac{\Al_{\plus}}{2 \sqrt{6}} & 0 & 0 & 0 & \frac{\Al_{\plus}}{10 \sqrt{6}} & 0 & \frac{\Al_{\minus}}{10} \\
 \frac{\Al_{\minus}}{2 \sqrt{5}} & 0 & 0 & 0 & \frac{\Al_{\minus}}{2 \sqrt{15}} & 0 & 0 & 0 \\
 0 & \frac{\Al_{\minus}}{2 \sqrt{10}} & 0 & \frac{\Al_{\plus}}{2 \sqrt{15}} & 0 & \frac{\Al_{\minus}}{2 \sqrt{10}} & 0 & 0
   \\
 \frac{\Al_{\plus}}{2 \sqrt{30}} & 0 & \frac{\Al_{\minus}}{2 \sqrt{30}} & 0 & \frac{\Al_{\plus}}{2 \sqrt{10}} & 0 &
   \frac{\Al_{\minus}}{2 \sqrt{10}} & 0 \\
 0 & \frac{\Al_{\plus}}{2 \sqrt{10}} & 0 & 0 & 0 & \frac{\Al_{\plus}}{2 \sqrt{10}} & 0 & \frac{\Al_{\minus}}{2 \sqrt{15}}
   \\
 0 & 0 & \frac{\Al_{\plus}}{2 \sqrt{5}} & 0 & 0 & 0 & \frac{\Al_{\plus}}{2 \sqrt{15}} & 0 \\
 0 & 0 & 0 & \frac{\Al_{\minus}}{\sqrt{10}} & 0 & 0 & 0 & 0 \\
 0 & 0 & 0 & 0 & \frac{\Al_{\minus}}{\sqrt{15}} & 0 & 0 & 0 \\
 0 & 0 & 0 & \frac{\Al_{\plus}}{5 \sqrt{6}} & 0 & \frac{\Al_{\minus}}{5} & 0 & 0 \\
 0 & 0 & 0 & 0 & \frac{\Al_{\plus}}{5 \sqrt{2}} & 0 & \frac{\Al_{\minus}}{5 \sqrt{2}} & 0 \\
 0 & 0 & 0 & 0 & 0 & \frac{\Al_{\plus}}{5} & 0 & \frac{\Al_{\minus}}{5 \sqrt{6}} \\
 0 & 0 & 0 & 0 & 0 & 0 & \frac{\Al_{\plus}}{\sqrt{15}} & 0 \\
 0 & 0 & 0 & 0 & 0 & 0 & 0 & \frac{\Al_{\plus}}{\sqrt{10}}
\end{array}
\right)
\end{equation}

To obtain the effective Hamiltonian, we follow Klein
\cite{Klein1974}. The notation of that work is somewhat obscure,
so for ease of understanding we repeat the main results.  From
Equation (A7) of that work, we have the $t$-order contribution to
the effective Hamiltonian
\begin{equation}
h_{\rm eff}^{(t)} = \sum_{\{k\}} A_{\{k\}} O_{\{k\}}
\end{equation}
Where $k_1, \ldots, k_{t-1}$ are non-negative integers, the $A$
are real coefficients, the $O$, denoted  ``$(k_1,k_2,\ldots,k_t)$''
by Klein, are operators, and the sum is taken over all ${\{k\}}$
satisfying $\sum_{l=1}^{t-1} k_l = t-1$. The  $A$ are given in
Table I of that work and the $O$ are given in Equation (A1) as
\begin{equation}
O_{\{k_1,\ldots,k_{t-1}\}} \equiv P_0 V R^{(k_1)} V R^{(k_2)}
\ldots V R^{(k_{t-1})} V P_0
\end{equation}
with $P_0$ being the projector onto the degenerate subspace and by
Equation (II.A.5)
\begin{equation}
R^{(k)} \equiv \left\{
\begin{array}{lr}
P_0 & k=0 \\
\left(\frac{1-P_0}{E_0-H^{(0)}} \right)^k & k>0
\end{array}
\right.
\end{equation}
where $E_0$ is the energy of the degenerate subspace.  In our case
we have chosen $E_0 = 0$.

We can then directly calculate the second- and fourth-order
contributions.  We are only concerned with $h_{\rm eff}$ as it
acts on the $F=1$ subspace, that is, with a $3\times 3$ matrix,
and it is convenient to express it in terms of the pseudo-spin
components $j_0,j_x,j_y,j_z$ and the Stokes components $S_0, S_x,
S_y, S_z$ defined above.  Summing the second-order
contributions we find $H_{\rm eff}^{(2)}$ of equation \ref{eq:HEff2}
With $B
\equiv - {D_{JJ'}^2}/{48 \delta_0 \delta_1 \delta_2 \hbar} $, 
 \begin{eqnarray}
\alpha^{(1)} & = &  B  ( 5 \delta_0 \delta_1-5 \delta_0 \delta_2-4
\delta_1 \delta_2
)\\
\alpha^{(2)} & = &  B  ( \delta_0 \delta_1-5 \delta_0 \delta_2+4
\delta_1 \delta_2)
\end{eqnarray}


Similarly, the fourth-order contribution is, dropping terms in
$S_0^2$,
\begin{eqnarray}
H_{\rm eff}^{(4)} &=& \beta_J^{(0)} S_Z^2 J_0 + \beta_N^{(0)}
S_Z^2 N_A +  \beta^{(1)}  S_0 S_Z J_Z  \nonumber \\ & & +
\beta^{(2)}  S_0 (S_X J_X + S_Y J_Y).
\end{eqnarray}
Note that the term in $N_A$ arises because $h_{\rm eff}^{(4)}$
contains a self-rotation term of the form $\beta_{m=0}^{(0)} S_Z^2
P_{m=0}$ where $P_{m=0}$ is a projector onto the state
$\ket{F=1,m_F=0}$. We express this in terms of $J_0$ and $N_A$
using $\sum_i P_{m=0}^{(i)} = \sum_i (I_3^{(i)}- j_{0}^{(i)}) =
N_A - J_0$.

\begin{figure}
\centering
{\includegraphics[width=.85\textwidth]{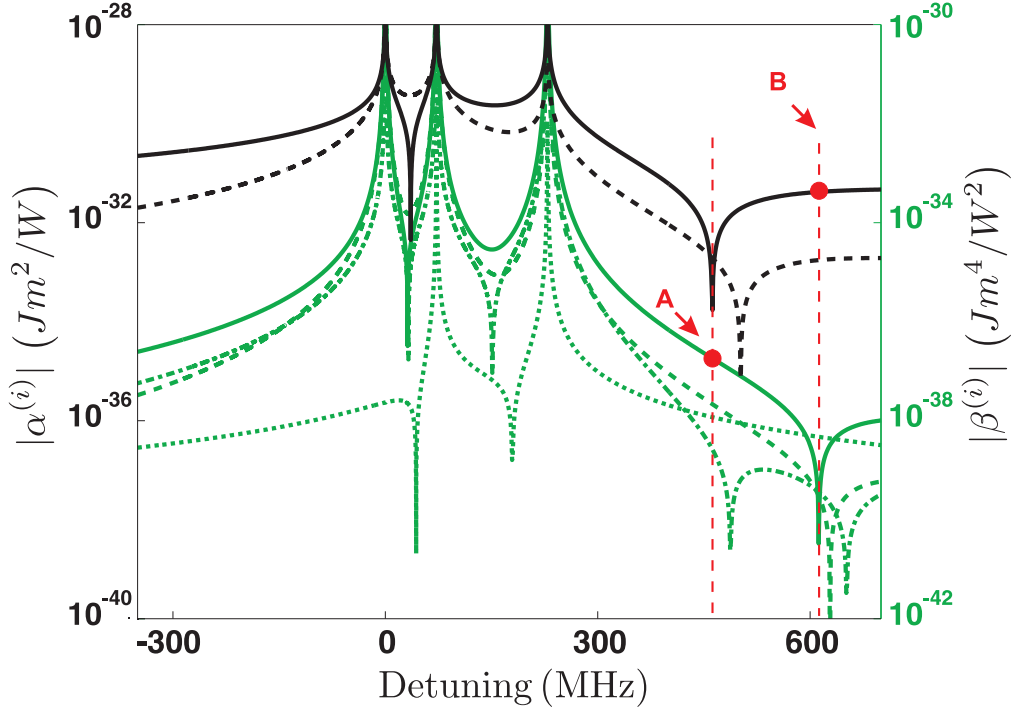}}

\caption{(color online). Spectra of the terms of the effective Hamiltonian. First two curves from the top, left axis: continuous, $\alpha^{(1)}$; dashed, $\alpha^{(2)}$. Lower curves, right axis: continuous, $\beta^{(1)}$; dotted, $\beta^{(0)}_N$; dashed, $\beta^{(2)}$; dot-dashed, $\beta^{(0)}_J$. Detuning [MHz] is relative to the transition $F=1\rightarrow F^\prime=0$ of ${}^{87}$Rb $D_2$ transition.  Points A, B, indicate detunings at which $\alpha^{(1)}$ or $\beta^{(1)}$ vanish.}
\label{fig:alphabeta}
\end{figure}

With $C \equiv{D_{JJ'}^4}/{1152 \delta_0^3 \delta_1^3 \delta_2^3
\HFSSym \hbar^3}$, the coefficients, shown graphically in  Figure
\ref{fig:alphabeta}, are
\begin{eqnarray}
 \beta_J^{(0)}& =& C
\left( 12 \delta_0^3 \delta_1^2 \delta_2^2-4 \delta_0^3 \delta_1
\delta_2^3+12 \delta_0^3 \delta_1^3 \HFSSym -10 \delta_0^3
\delta_1^2 \delta_2 \HFSSym-12 \delta_0^2 \delta_1^3 \delta_2
\HFSSym \right. \nonumber \\ & & \left. -10 \delta_0^3 \delta_1 \delta_2^2 \HFSSym -12 \delta_0
\delta_1^3 \delta_2^2 \HFSSym+20 \delta_0^2 \delta_1 \delta_2^3
\HFSSym +20 \delta_0 \delta_1^2
\delta_2^3 \HFSSym
\right)\\
\beta_N^{(0)} & =& C
\left( -12 \delta_0^3 \delta_1^3 \delta_2-24 \delta_0^3 \delta_1^2
\delta_2^2+4 \delta_0^3 \delta_1 \delta_2^3
\right) \\
 \beta^{(1)}  & =& C
\left( -9 \delta_0^3 \delta_1^3 \delta_2+6 \delta_0^3 \delta_1^2
\delta_2^2+3 \delta_0^3 \delta_1 \delta_2^3+35 \delta_0^3
\delta_1^3 \HFSSym-5 \delta_0^3 \delta_1^2 \delta_2 \HFSSym \right. \nonumber \\ & & \left.-4
\delta_0^2 \delta_1^3 \delta_2 \HFSSym-5 \delta_0^3 \delta_1
\delta_2^2 \HFSSym-4 \delta_0 \delta_1^3 \delta_2^2 \HFSSym-25
\delta_0^3 \delta_2^3 \HFSSym -20
\delta_0^2 \delta_1 \delta_2^3 \HFSSym \right. \nonumber \\ & & \left.-20 \delta_0 \delta_1^2
\delta_2^3 \HFSSym-16 \delta_1^3 \delta_2^3 \HFSSym
\right) \\
 \beta^{(2)}  & =&  C
\left( 3 \delta_0^3 \delta_1^3 \delta_2-6 \delta_0^3 \delta_1^2
\delta_2^2+3 \delta_0^3 \delta_1 \delta_2^3+7 \delta_0^3
\delta_1^3 \HFSSym-15 \delta_0^3 \delta_1^2 \delta_2 \HFSSym\right. \nonumber \\ & & \left.+16
\delta_0^2 \delta_1^3 \delta_2 \HFSSym-15 \delta_0^3 \delta_1
\delta_2^2 \HFSSym+16 \delta_0 \delta_1^3 \delta_2^2 \HFSSym-25
\delta_0^3 \delta_2^3 \HFSSym\right. \nonumber \\ & & \left.+16
\delta_1^3 \delta_2^3 \HFSSym \right)
 \end{eqnarray}


\section{Application to nonlinear metrology}
The $\beta$ terms are nonlinear in $\bf S$, indicating a photon-photon interaction. We expect these terms
to describe polarization effects of fast electronic nonlinearities including
saturation and four-wave mixing.  As in the linear case, the frequency dependence of the $\beta$ terms provides considerable 
flexibility in designing a light-matter interaction.   Applied to quantum metrology, these terms produce SH scaling,
because they are nonlinear in the $S$ collective variables, while the
atomic variables $J_i$, $N_A$ play the role of the parameter.  
The $\beta^{(0)}$ and $\beta^{(1)}$
terms are analogous to Hamiltonians considered by Boixo {\it{et al.}} \cite{Boixo2007}.
The $\beta^{(1)}$ term $\propto S_0 S_z$, in particular, achieves SH scaling
without input or generated
entanglement \cite{Boixo2008PRL}.   The $\beta^{(2)}$ terms describes a
nonlinear tensorial contribution, and does not appear to
have been considered yet for nonlinear metrology.


\section{Quantum Noise}
To understand the
quantum noise in this system, we define polarization operators $\PE_i \equiv
\mtext{\frac{1}{2}} (a_+^\dagger,a_-^\dagger)\sigma_i
(a_+,a_-)^{\rm{T}} $, where $a_\pm$ are annihilation operators for
the $\pm$ circular polarizations of a mode defined by the pulse
shape. These obey angular momentum commutation relations
$[\PE_i,\PE_j] = i \varepsilon_{ijk} \PE_k$ and are related to the
Stokes parameters by $\PE_i = S_i/\mtext{2}\gamma$, where $\gamma
\equiv {\hbar \omega Z_0 }/{{2}T A}$ is the single-photon
intensity, $T$ is the pulse duration, $A$ is the
beam area, and $Z_0$ is the impedance of free space. The
total number of photons is $N_L = \mtext{2}\PE_0$. For a typical input,
a coherent state, $\myexpect{(\PE_X,\PE_Y,\PE_Z)} =
(\mtext{\PE_0},0,0)$ and $\var(\PE_i) = \mtext{\PE_0/2}$.

Evolution under this effective Hamiltonian produces, to first
order in the interaction time $\tau$, \be\label{eq:in_out}
\PE_Y\upout = \PE_Y\upin +\frac{\tau}{\hbar} (\alpha^{(1)} +
\beta^{(1)}\gamma \PE_0) \gamma\PE_X\upin J_Z\upin\ee
plus terms containing $\PE_Z\upin J_X\upin$ that are negligible for the given input 
coherent state of the light. This evolution physically corresponds to a paramagnetic 
Faraday rotation of the input linear polarization. In a metrological scheme one would
measure this polarization rotation and from it estimate
the atomic variable $J_Z$.

For small rotation, i.e. $\phi \equiv \PE_Y\upout/ \PE_X\upin = 
\gamma \tau \hbar^{-1} (\alpha^{(1)} + \beta^{(1)} \gamma \PE_0) J_Z\upin \ll
1$, we note that the input polarization noise dominates: $
\var(\PE_Y\upout) = \var(\PE_Y\upin ) + \phi^2 \var(\PE_X) \approx
\var(\PE_Y\upin ) = \mtext{\PE_0/2}$, and that the signal-to-noise ratio equals one when
$\myexpect{\PE_Y\upout}^2 = \var(\PE_Y\upin),$ i.e., when \be\label{eq:SNR1}
\frac{\tau^2 \gamma^2}{\hbar^2} \otext{N_L^2} \mtext{\PE_0^2} (\alpha^{(1)} +
\beta^{(1)}\gamma \PE_0)^2 (J_Z\upin )^2
= \otext{{N_L}}\frac{\PE_0}{2}.\ee

We can identify the value of $J_Z\upin$ that solves Eq.
(\ref{eq:SNR1}) as the sensitivity, or precision of the estimation, $\delta J_Z$. We find \be\label{eq:sens_th} \delta J_Z = \hbar |\otext{\sqrt{2}}\tau\gamma(\alpha^{(1)}\ntext{N_L^{1/2}}\otext{\PE_0^{1/2}} + \frac{
\beta^{(1)}\gamma}{\mtext{2}}\ntext{N_L^{3/2}}\otext{\PE_0^{3/2}})|^{-1}.\ee Thus the
sensitivity will have a transition from shot-noise to SH scaling
with increasing $N_L$.  As indicated in Figure \ref{fig:alphabeta}, there are points in the spectrum where either 
$\alpha^{(1)}$ or $\beta^{(1)}$ vanish, allowing pure nonlinear or pure linear estimation of the same atomic variable. 

In another scenario, an unpolarized input state $\myexpect{(\PE_X,\PE_Y,\PE_Z)} =
(0,0,0)$ gives rise to dynamics dominated by the $\beta^{(0)}$ terms $\propto S_Z^2$, sometimes called the
``one-axis twisting Hamiltonian.''  This describes a self-rotation of the optical polarization, and can
be used to generate polarization squeezing and also to obtain sensitivity scaling as $N_L^{-3/2}$ in the estimation of
$\beta^{(0)}_J J_0 + \beta^{(0)}_N N_A$, using an entanglement-generating strategy described in 
reference \cite{Boixo2008PRL}.

%

\section{Conclusion} We have generalized the formalism of continuous collective variables to the nonlinear regime.  The resulting nonlinear effective Hamiltonian includes several distinct nonlinear couplings with strengths widely tunable via the probe light frequency.  This allows the production of model Hamiltonians proposed for nonlinear metrology, including both models that generate entanglement and those which achieve super-Heisenberg scaling without entanglement.   Similar nonlinear probing techniques could improve optical probing of atomic clocks \cite{Appel2009, SchleierSmith2010} and atomic magnetometers \cite{RomalisAPL2006, RomalisNP2007, WasilewskiPolzik2010}. 
Unlike previous proposals, the atomic ensemble system allows both linear and nonlinear estimation of the same atomic variables.

\section*{Acknowledgement}
We thank M. Koschorreck, B. Dubost, N. Behbood, A. Cer\`e and R. Sewell for helpful discussion. F. Illuminati contributed essential ideas at an early stage. This work was funded by the Spanish Ministry of Science and Innovation under the FPU project and the Consolider-Ingenio 2010 Project ``QOIT''

\section*{References}

\bibliographystyle{unsrt}


\end{document}